\documentclass[twocolumn]{aastex631}

\usepackage{amsmath,amssymb}
\usepackage{booktabs}
\usepackage{multirow}
\usepackage{array}
\usepackage{xcolor}
\usepackage[switch,mathlines]{lineno}

\newcommand{\mures}{\ensuremath{\Delta\mu}}

\newcommand{\lgbm}{LightGBM}
\newcommand{\lsst}{LSST}

\newcommand{\bbc}{BBC}
\newcommand{\salt}{SALT3}

\newcommand{\logm}{\ensuremath{\log_{10}(M_\star/M_\odot)}}

\shorttitle{ML Closure Audits for LSST Simulations}
\shortauthors{Mitra et al.}

\begin{document}
\title{Machine Learning Closure Audits for \lsst\ Photometric Supernova Cosmology}

\author[0000-0000-0000-0000]{Ayan Mitra}
\affiliation{Center for AstroPhysical Surveys, National Center for Supercomputing Applications, Urbana, IL 61801, USA}
\affiliation{Department of Astronomy, University of Illinois at Urbana Champaign, Urbana, IL 61801, USA}

\correspondingauthor{Ayan Mitra}
\email{ayan@illinois.edu}

\begin{abstract}
Modern and next generation supernova cosmology analyses rely on end to end simulations to train photometric classifiers, characterise selection effects, validate light curve models, and calibrate distance bias corrections. Standard closure tests based on global Hubble diagram summaries can miss multivariate structure that remains in post correction residuals. We introduce a supervised machine learning closure audit that tests whether measured observables can predict the bias corrected Hubble residuals $\Delta\mu$. We apply the audit to LSST Type Ia supernova simulations from two independent analyses: the M23 mock data sets of \citet{Mitra2023}, with spectroscopic redshift and photometric redshift samples, and the predominantly photometric LSST like M25 simulation of \citet{Mitra2025}. Standard one dimensional Redshift binned diagnostics explain $<1\%$ of the residual variance ($R^2 < 0.01$), suggesting apparent closure. In contrast, out of fold LightGBM models recover up to $98.2\%$ of the variance in the simulated residuals ($R^2 = 0.982$), revealing structured residual predictability. Applying the same audit directly to the real Dark Energy Survey 5 Year (DES 5YR) spectroscopic sample yields a held out $R^2 = 0.725$. SHAP feature attribution rankings are highly consistent between independently trained M25 and DES models (Spearman $\rho = 0.802$), with apparent magnitudes, signal to noise ratios, and redshift dominating the shared predictive hierarchy. The resulting scorecard provides a diagnostic protocol for comparing mock ensembles and real observations, identifying residual non closure before cosmological parameters are unblinded.
\end{abstract}


\section{Introduction}
\label{sec:intro}

Type Ia supernovae (SNe~Ia) are crucial probes of cosmic acceleration \citep{Phillips1993,Riess1998,Perlmutter1999}. Modern surveys such as the Vera C.\ Rubin Observatory \lsst\ will gather orders of magnitude more photometrically classified SNe~Ia than current samples, shifting the limiting uncertainty from statistical precision to systematic errors in calibration, selection, and survey modelling \citep{LSSTDESC2018}.

In standard analyses, light curves are fitted (e.g., with \salt; \citealt{Kenworthy2021}) and corrected for selection effects via simulation based frameworks like BEAMS with Bias Corrections (\bbc; \citealt{Kessler2017}). The resulting Hubble residuals $\mures \equiv \mu -  \mu_{\rm model}$, where $\mu_{\rm model}$ is computed under the fiducial cosmology specified by the parent analysis, should ideally be featureless. However, residuals can retain structures from selection correction errors, redshift uncertainties, contamination, or unmodelled environmental factors.

Standard closure tests evaluate whether the pipeline recovers the input cosmology globally, but can overlook localised, multivariate failures that average out over the full sample. We use simulations as controlled environments to inspect what systematic residual structures persist after corrections. Host galaxy environment is a key example of residual systematics. In standard analyses, a binary ``mass step'' corrects for the correlation between standardised brightness and host galaxy stellar mass \citep{Kelly2010,Sullivan2010}. However, the step amplitude depends on selection corrections, and mass may act as a proxy for complex dust distributions and population properties \citep{Brout2021,Smith2020}.

The goal of this paper is to  ask the question: do residuals that look closed in one dimensional redshift summaries remain predictable when a ML model uses the measured light curve, host galaxy, and survey quality information? We answer this question first on the M23 and M25 mock catalogues, then on the Dark Energy Survey 5 Year (DES 5YR) sample.
The motivation for this analysis is that supervised machine learning can detect complex, multivariate residual structure without pre-specifying a functional form. In this paper, predictability is a diagnostic, not a proposed distance correction: high predictability can arise partly because $\mures$ is constructed from fitted light curve and redshift parameters, but it also shows where the corrected residual field is structured. The two main metrics are the out of fold coefficient of determination, \(R^2\), and the root mean square prediction error, RMSE. Both are computed only on held out events. \(R^2\) compares the model prediction with a constant mean residual prediction: \(R^2=0\) means no improvement over that mean, and \(R^2=1\) means perfect prediction. RMSE is the typical prediction error in magnitudes. For a closure audit, lower \(R^2\) is the desired physical outcome; a larger \(R^2\) means that more residual structure is present and measurable. To interpret the model, we compute SHAP (SHapley Additive exPlanations) values \citep{Lundberg2017}. SHAP values  rank which observables contribute most to the prediction. A larger SHAP value means stronger predictive attribution in the fitted model; it however neither implies better or worse cosmological result nor does it prove causation. Section~\ref{sec:metrics} gives the explicit definitions.

The main findings of this work are:
\begin{enumerate}
  \item \textbf{Residuals are predictable even when redshift summaries look closed.} Redshift binned mean residuals explain less than 1\% of the variance, but the full audit models explain up to 98.2\% in the simulations and 72.5\% in DES 5YR.
  \item \textbf{Brightness and measurement quality carry most of the signal.} Apparent magnitudes and signal to noise measurements dominate the feature tests. They locate events relative to detection and selection limits, rather than pointing to a unique problem in one passband.
  \item \textbf{M25 and DES use similar shared predictors.} Independent models trained separately on M25 and DES 5YR have a Spearman rank correlation of $\rho = 0.802$ between their SHAP feature rankings. Even when restricted to observables common to both data products, the variables that predict residuals are ranked similarly.
\end{enumerate}
 It can be applied to any pipeline that provides residuals and measured observables however with the disclaimer that  these results motivate the audit as a diagnostic layer rather than a replacement distance estimator.

\section{Data}
\label{sec:data}
\subsection{M25 LSST Photometric Simulation}
\label{sec:m25}

Our primary data set is the M25 simulation, representing the photometrically classified SN~Ia sample with photometric redshift information from \citet{Mitra2025}. M25 is built using the Extended LSST Astronomical Time series Classification Challenge (ELAsTiCC) programme and host galaxy framework \citep{Narayan2023ELaSTiCC}. It combines a high redshift Deep Drilling Field (DDF) component ($\sim\!87\%$) using host galaxy photometric redshifts and the SCONE photometric classification algorithm \citep{Qu2021} with a low redshift Wide Fast Deep (WFD) spectroscopic anchor sample ($\sim\!13\%$). The simulation contains 25 independent realisations, each with roughly \(6{,}000\) bias corrected events after analysis cuts. In this paper we analyse the total simulated Type Ia subset from the 25 realizations: the 124,132 events whose known simulated input type is SN~Ia. This removes known non Ia contaminants so that the audit isolates residual structure in the Type Ia cosmology sample after the parent M25 light curve fitting, selection, and bias correction pipeline.

M25 light curves are fitted with SALT3 \citep{Kenworthy2021}. Its intrinsic scatter follows the BS21 dust based model \citep{Brout2021}, introducing host mass dependent dust distributions. Crucially, the nominal standardisation is ``mass blind'' ($\gamma=0$), meaning the bias correction grid is host mass independent. Because the simulation includes mass dependent dust scatter while the correction is mass blind, a residual host mass step is expected to emerge. This allows us to inspect whether host mass dependent residual structure reappears after correction.

\subsection{M23 and DES 5 Year Benchmark Data Sets}
\label{sec:m23}
\label{sec:simulation_roles}

We also analyse the M23 simulation \citep{Mitra2023}, which provides independent spectroscopic redshift (55,279 events) and photometric redshift (145,429 events) data sets across 25 realisations. M23 light curves are fitted with SALT2 \citep{Guy2007,Guy2010}. M23 uses the G10 intrinsic scatter model \citep{Guy2010} and a mass blind bias correction with $\gamma=0$, and it does not include photometric classification variables. Table~\ref{tab:pipeline_comparison} summarises these configurations.

As an independent real data companion audit, we apply the same scorecard within the Dark Energy Survey 5 Year (DES 5YR) spectroscopic SN~Ia sample from the DES SN5YR analysis \citep{Vincenzi2024}. DES 5YR also uses SALT3 light curve fitting. The feature level product used here contains 1,820 SNe~Ia after matching the available DES columns to our audit requirements. No model is trained on M25 and applied to DES; the DES model is trained and cross validated within DES, then compared with M25 only at the level of shared feature rankings.

It is important to separate the roles of the different simulated products. The M23/M25 data sets audited in this paper are mock data realisations: simulated catalogues processed as if they were observed data. The parent analyses also use other simulations as calibration inputs, including the bias correction simulations used by \bbc, classifier training simulations for SCONE in M25, and survey observing simulations that encode cadence and selection. Those calibration simulations are not separate audit rows here; they are part of the pipeline that produced the residuals being audited. Future real LSST observations will replace the mock data realisation role, while classifier training and bias correction simulations will remain calibration inputs. The real data scorecard should therefore be compared with a matched mock ensemble scorecard.

\begin{deluxetable*}{lllll}
\tablecaption{Analysis properties relevant to the independent closure cases. The columns are: (1) \textbf{Data set}: the specific data set analysed; (2) \textbf{Redshift/classification}: the redshift source and, where applicable, the photometric classifier; (3) \textbf{Light curve fit}: the light curve model used for distance standardisation; (4) \textbf{Host mass audit}: whether host galaxy stellar mass is available for host step diagnostics; and (5) \textbf{Nominal bias correction}: the simulated intrinsic scatter model and distance bias correction configuration. M23 spec-\(z\), M23 photo-\(z\), and M25 are independent pipeline configurations with several simultaneous differences, including light curve model, scatter model, sample construction, available outputs, and bias correction dimensions. They must not be interpreted as a controlled sequence in which only one effect changes.\label{tab:pipeline_comparison}}
\tabletypesize{\scriptsize}
\tablewidth{0pt}
\tablehead{
  \colhead{Data set} & \colhead{Redshift/classification} &
  \colhead{Light curve fit} & \colhead{Host mass audit} & \colhead{Nominal bias correction}
}
\startdata
M23 spec-\(z\) & spec-\(z\); no photometric classifier & SALT2 & - &
G10 scatter; mass blind; \(\gamma=0\) fixed \\
M23 photo-\(z\) & host photo-\(z\); no photometric classifier  & SALT2 & - &
G10 scatter; mass blind; \(\gamma=0\) fixed \\
M25  & host photo-\(z\) DDF + spec-\(z\) anchor; SCONE & SALT3 & available &
BS21 dust; mass blind; \(\gamma=0\) fixed \\
\enddata
\end{deluxetable*}

\section{Methods}
\label{sec:methods}

\subsection{Closure Audit Design and Metrics}
\label{sec:features}
\label{sec:metrics}

The M25 catalogue contains 76 candidate observables after excluding event identifiers and simulation truth quantities. We compare increasingly informative predictors: (1) a one dimensional redshift binned mean, (2) a four feature \emph{Tripp standardisation set} \(\{x_1,c,m_B,z_{\rm HD}\}\), and (3) a full 76 feature audit model (or 58 features for M23). The full M25, M23, and DES shared feature inventories are listed in Appendix~\ref{sec:appendix_feature_inventory}. We call the four feature set the Tripp ``core'' because these variables enter the standard SN~Ia distance relation directly; the term does not imply that they are the only physically important quantities. We evaluate all predictors on held out mock data realisations using the coefficient of determination $R^2$ and root mean square error (RMSE) in magnitudes:
\begin{equation}
\begin{aligned}
 R^2 &= 1  - \frac{\sum_i(\mures_i- \widehat{\mures}_i)^2}{\sum_i(\mures_i -\overline{\mures})^2}, \\
 {\rm RMSE} &= \sqrt{\frac{1}{N}\sum_i(\mures_i- \widehat{\mures}_i)^2},
\end{aligned}
\end{equation}
where $\widehat{\mures}_i$ is the out of fold prediction and $\overline{\mures}$ is the target residual mean. An $R^2$ of 1 indicates perfect prediction, while $R^2 \approx 0$ indicates performance no better than a constant mean residual. In a closure audit, the desired physical outcome is low predictability; where as a high $R^2$  exposes residual structure that the standard corrections did not remove. Because the target residual \mures\ is mathematically constructed from fitted light curve parameters and redshift (Equation~\ref{eq:standardization}), the Tripp set model has an expected arithmetic component. We therefore treat its performance, \(R^2_{\rm core}\), as an arithmetic baseline rather than as a standalone claim of physical failure. The diagnostic quantities are (1) how much residual structure remains predictable in this basic standardisation coordinate system and (2) the incremental gain \(\Delta R^2 = R^2_{\rm full}-R^2_{\rm core}\) obtained by adding observables that are not direct terms in the Tripp relation.

\subsection{Feature Groups and Leakage Checks}
\label{sec:feature_groups}

To interpret residual predictability and prevent circular reuse of information, we classify candidate features into four groups based on their mathematical and physical relationship to the distance estimation pipeline:
\begin{enumerate}
    \item \textbf{Target adjacent standardisation features:} light curve fit and redshift parameters (\(x_1, c, m_B, z_{\rm HD}\)) that directly construct the target residual \mures\ via Equation~\ref{eq:standardization}. Predictability from this set defines \(R^2_{\rm core}\) \label{cut_1}. 
    \item \textbf{Physical environment diagnostics:} astrophysical properties of the host galaxy, such as stellar mass \logm, star formation rate, rest frame colours, and Milky Way reddening \(E(B-V)\). They  can trace unmodelled dust or stellar population systematics.
    \item \textbf{Fit quality and selection limit tracers:} Parameters such as fit \(\chi^2\), fit probability, and band specific signal to noise ratios. These do not enter the distance calculation directly, but they locate each event relative to the survey's detection limits and selection cuts.
    \item \textbf{Analysis stage outputs:} intermediate quantities produced by the bias correction and classification pipeline, such as per object distance bias corrections, distance uncertainty rescalings, and classification probabilities. These can carry information about the simulation selection grid. 
\end{enumerate}

\subsection{Model Training and Feature Attribution}
\label{sec:lgbm}
\label{sec:shap}
\label{sec:standardization}

We train LightGBM \citep{Ke2017} using fixed baseline hyperparameters ($N_{\rm trees}=1000$, learning rate $= 0.05$, $N_{\rm leaves}=63$) for the predictive performance metrics. Cross validation uses a simulation held out five fold scheme, where the 25 realisations are split into five independent groups to eliminate leakage between training and testing. Additionally we compare LightGBM against a linear Ridge regression ($\alpha=1$) and an ExtraTrees regressor on identical folds. SHAP only diagnostic runs use the same feature sets and folds but a 600 tree LightGBM model and a capped SHAP evaluation subset for computational tractability; the fold stability check in Section~\ref{sec:shap_results} verifies that the resulting rankings are stable.  For each supernova, each feature receives a signed contribution, in mag, to the predicted Hubble residual $\mures$, defined under the standardised Tripp relation:
\begin{equation}
  \label{eq:standardization}
  \mures = m_B + \alpha x_1 - \beta c + M_0 + \Delta\mu_{\rm bias} - \mu_{\rm model}(z),
\end{equation}
where $\Delta\mu_{\rm bias}$ is the simulation derived distance bias correction and $\mu_{\rm model}(z)$ is the distance modulus under the reference cosmology. Feature importances are evaluated by averaging absolute SHAP values across folds.

\section{Results}
\label{sec:results}

\subsection{Independent Closure Audit Scorecard}
\label{sec:closure_scorecard}

\begin{figure*}[p]
\centering
\includegraphics[width=0.96\textwidth]{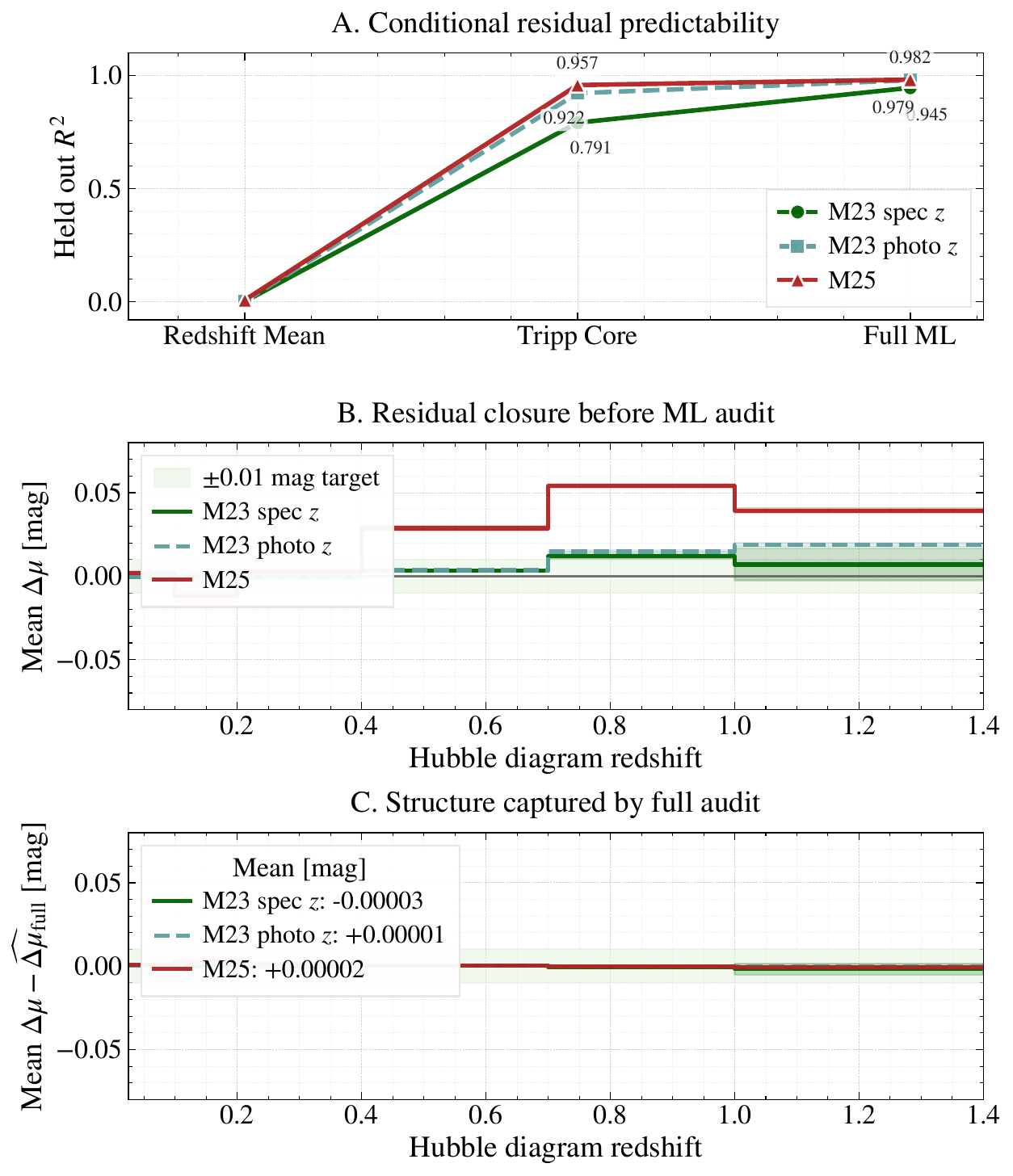}
\caption{Independent closure audit metrics. Panel A: held out predictive performance for increasingly informative baselines and models. Panel B: mean raw residual \(\langle\Delta\mu\rangle\) in six redshift bins; these are the binned residual means measured from the parent M23 and M25 mock data products. Panel C: mean residual after subtracting the full out of fold machine learning prediction, \(\langle\Delta\mu -\widehat{\Delta\mu}_{\rm full}\rangle\), in the same redshift bins. The Panel C legend reports the weighted mean of the plotted post audit residual bins in mag. Panels B and C do not show \(R^2\); instead they show binned residual means in mag. The scorecard exposes structure that is nearly invisible to redshift bin mean baselines.\label{fig:lsst_closure_audit}}
\end{figure*}

We begin by applying the closure audit to three independent simulated data sets: M23 spectroscopic $z$, M23 photometric $z$, and M25. The results are summarised in the independent closure audit scorecard (Table~\ref{tab:lsst_closure_audit}; Figure~\ref{fig:lsst_closure_audit}).

As a null baseline, we compute a one dimensional redshift binned mean predictor, \(R^2_z\). For each validation fold, we divide the training sample into six redshift bins, calculate the mean Hubble residual in each training bin, and assign the corresponding training bin mean to held out events. This measures how much residual structure can be captured by a simple redshift dependent correction. The \(R^2_z\) column shows that this baseline explains less than 1\% of the variance in all three simulated data sets.

Next, we evaluate two additional machine learning models:
\begin{enumerate}
  \item The Tripp standardisation model ($R^2_{\rm core}$) (Sec. \ref{sec:feature_groups} cut \ref{cut_1}), which is trained using only stretch ($x_1$), colour ($c$), apparent magnitude ($m_B$), and redshift ($z_{\rm HD}$). This model captures how much residual structure is predictable from the core variables of SN~Ia standardisation. It reaches $R^2_{\rm core} = 0.791$, $0.922$, and $0.957$ in the three data sets, respectively.
  \item The full audit model ($R^2_{\rm full}$), which uses all available observables and analysis stage outputs (58 features for M23, 76 for M25). This model captures more complex multivariate residual relationships, reaching $R^2_{\rm full} = 0.945$, $0.979$, and $0.982$, respectively.
\end{enumerate}

On the scale defined in Section~\ref{sec:metrics}, the full models reduce squared prediction error relative to a constant mean reference by \(94.5\%\), \(97.9\%\), and \(98.2\%\). More diagnostically, the gain from the Tripp standardisation model to the full model is \(\Delta R^2=0.154\), \(0.057\), and \(0.024\). 

\par\smallskip
\noindent\begin{minipage}{\columnwidth}
\refstepcounter{table}\label{tab:lsst_closure_audit}
\noindent{\footnotesize\textbf{Table~\thetable.} Independent closure audit scorecard for LSST mock data sets.\par}
\vspace{2pt}
\centering
\scriptsize
\resizebox{\columnwidth}{!}{%
\begin{tabular}{@{}lrrrrrr@{}}
\toprule
Data set & \(N\) & \(\langle\Delta\mu\rangle\) & \(\sigma\) & \(R^2_z\) & \(R^2_{\rm core}\) & \(R^2_{\rm full}\) \\
\midrule
M23 spec $z$ & 55,279 & +0.003 & 0.116 & 0.001 & 0.791 & 0.945 \\
M23 photo $z$ & 145,429 & +0.010 & 0.150 & 0.002 & 0.922 & 0.979 \\
M25 & 124,132 & +0.035 & 0.225 & 0.008 & 0.957 & 0.982 \\
\bottomrule
\end{tabular}%
}
\vspace{2pt}
\noindent{\scriptsize Note---Each row is trained and evaluated using independent realisations from the listed mock-data set. \(R^2_z\) is the redshift-binned mean baseline, \(R^2_{\rm core}\) uses only \((x_1,c,m_B,z_{\rm HD})\), and \(R^2_{\rm full}\) uses the full audit model.\par}
\end{minipage}
\par\smallskip

The scorecard is the reusable product of this work.  It reports not only a
predictive metric, but also the null baselines, the gain from core fitted
variables to the full analysis outputs, the redshift dependent conditional
means, and the safeguards required to interpret them. Figure~\ref{fig:lsst_feature_gain} separates the leading full audit feature
gains from the closure metrics to make
clear that they are predictive attribution rather than scorecard acceptance
criteria.  The remainder of the paper focuses on M25 because its richer outputs
permit a detailed diagnosis of the detected structure.

\subsection{M25 Hubble Residual Predictability}
\label{sec:r2}

Table~\ref{tab:predictability_performance} summarises the M25 null baselines together with the out of fold cross validation performance of the M25 and DES 5YR audit models, the M25 reduced shared feature model, and two algorithmic checks: ExtraTrees and Ridge regression.

\begin{table*}[t]
\centering
\caption{Closure audit predictability and null baseline performance. Rows are grouped by sample and purpose. The first group gives M25 baselines: a held out constant mean residual, a six bin redshift mean, and a redshift plus host mass binned mean. The second group gives M25 audit models: the primary LightGBM model using all 76 M25 features, the LightGBM model using the 13 shared M25 and DES features, and two algorithmic checks using the same 76 M25 features as the primary model. The final group gives the DES 5YR feature audit. The columns are: (1) \textbf{Dataset / Model}: the sample and model configuration; (2) \textbf{${N_{\rm SNe}}$}: the total number of supernovae; (3) \textbf{${N_{\rm features}}$}: the number of input observables or model features, with the baseline rows counting the binned inputs used; (4) \textbf{${R^2}$}: the out of fold coefficient of determination relative to a constant mean residual; and (5) \textbf{RMSE (mag)}: the out of fold root mean squared prediction error. In a closure audit, low $R^2$ is the desired physical outcome. High $R^2$ means that the residual field is predictable and therefore not featureless.\label{tab:predictability_performance}}
\begin{tabular}{lcccc}
\hline\hline
Dataset / Model & $N_{\rm SNe}$ & $N_{\rm features}$ & $R^2$ & RMSE (mag) \\
\hline
\textbf{M25 simulation: baselines} \\
~~Held out constant mean residual       & 124,132 & 0 & 0.000 & 0.225 \\
~~Held out six redshift bin means       & 124,132 & 1 & 0.008 & 0.224 \\
~~Held out redshift bin + host mass means & 124,132 & 2 & 0.016 & 0.223 \\
\hline
\textbf{M25 simulation: audit models} \\
~~LightGBM, all M25 features            & 124,132 & 76 & 0.982 & 0.030 \\
~~LightGBM, shared M25 and DES features & 124,132 & 13 & 0.967 & 0.041 \\
~~ExtraTrees, all M25 features         & 124,132 & 76 & 0.887 & 0.075 \\
~~Ridge, all M25 features              & 124,132 & 76 & 0.857 & 0.085 \\
\hline
\textbf{DES 5YR: shared feature audit} \\
~~LightGBM, shared M25 and DES features & 1,820   & 13 & 0.725 & 0.139 \\
\hline
\end{tabular}
\end{table*}

The M25 \lgbm\ model achieves $R^2 = 0.982$ and RMSE \(=0.030\)~mag. Because \mures\ is constructed from light curve and redshift quantities, this high absolute predictability is partly expected and is not by itself the scientific result. Relative to the four feature Tripp standardisation model (\(R^2=0.957\)), the full audit gains \(\Delta R^2=0.024\), which corresponds to \(57\%\) of the squared error left by the Tripp model. The diagnostic value is therefore not the absolute \(R^2\) alone, but the repeatable conditional structures that remain stable under feature removal, null tests, and systematic variations. The all object performance is even higher ($R^2 = 0.997$). 
Table~\ref{tab:predictability_performance} also shows what the high \(R^2\) is not measuring. The raw M25 residual target has \(\sigma_{\mures}=0.225\)~mag and a mean offset of \(+0.035\)~mag. A held out constant mean has essentially zero explanatory power; a redshift binned mean explains only 0.8\% of the variance; and a redshift plus host binned mean explains only 1.6\%. These percentages are the corresponding \(R^2\) values in Table~\ref{tab:predictability_performance}. Thus the \lgbm\ RMSE of \(0.030\)~mag is not driven by a global offset, by \(\langle\mures(z)\rangle\), or by a coarse host redshift step alone. It reflects a structured residual field in the corrected simulation outputs.

The algorithm checks show that this conclusion is not unique to LightGBM. Ridge regression, a linear model using the same 76 M25 features, reaches \(R^2=0.857\) and RMSE \(=0.085\)~mag. ExtraTrees, an independent tree ensemble regressor using the same 76 M25 features, reaches \(R^2=0.887\), or \(R^2=0.884\) after removing the bias correction output feature group. Both alternatives are weaker than LightGBM, so we retain LightGBM as the precision diagnostic model, but they confirm that the residual field is broadly structured rather than an artefact of a single algorithm.

\subsection{Feature Importances}
\label{sec:shap_results}

\begin{figure*}[!tp]
\centering
\includegraphics[width=0.82\textwidth]{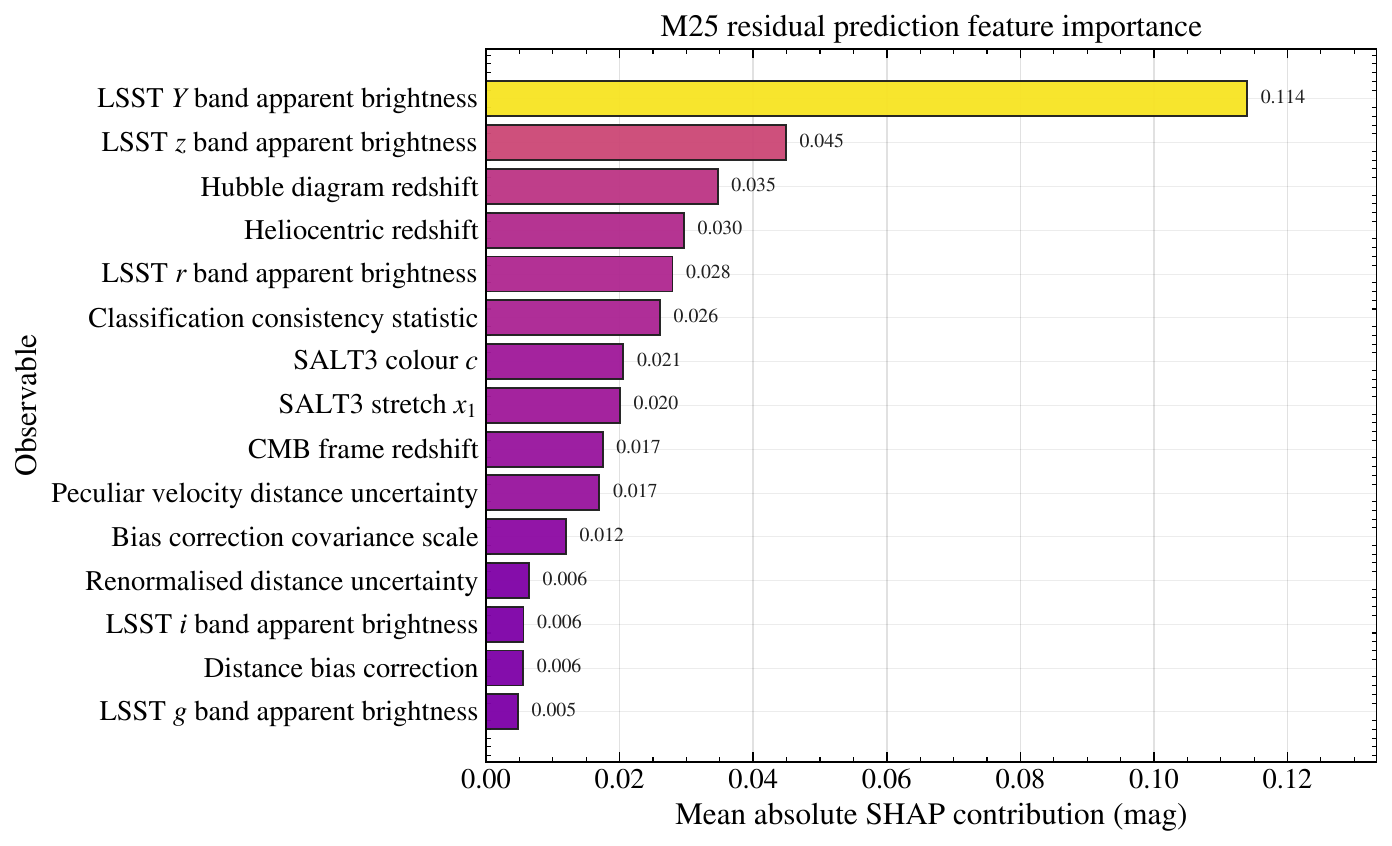}
\caption{SHAP feature importance for the M25 audit model. A larger bar means that the observable typically contributes more strongly to the model's predicted Hubble residual. The x axis is mean absolute SHAP contribution in mag. These are predictive attributions, not a ranking of physical causes or calibration error impacts.\label{fig:shap_beeswarm}}
\end{figure*}

Figure~\ref{fig:shap_beeswarm} shows the top 15 SHAP feature importance summary for the full M25 audit model, also  listed in Table~\ref{tab:top_shap}.

\begin{deluxetable*}{rllr}
\tablecaption{Top 15 M25 SHAP feature importances for the full audit model. The columns are: (1) \textbf{Rank}: the feature importance ordering from 1 to 15; (2) \textbf{Observable}: the input feature; (3) \textbf{Notes}: context for the feature; and (4) \textbf{Mean ${|\mathrm{SHAP}|}$ (mag)}: the average absolute contribution of the feature to the predicted distance residual. A larger value means stronger predictive attribution in this fitted model. It does not by itself imply a physical cause. These SHAP values use 600 boosting trees and a representative 20,000 event SHAP evaluation subset for computational tractability.\label{tab:top_shap}}
\tabletypesize{\footnotesize}
\tablewidth{0pt}
\tablehead{
  \colhead{Rank} & \colhead{Observable} & \colhead{Notes} &
  \colhead{Mean $|\mathrm{SHAP}|$ (mag)}
}
\startdata
 1 & $Y$ band apparent magnitude    & LSST band brightness; Pearson \(r=0.806\) with \(z_{\rm HD}\) & 0.114 \\
 2 & $z$ band apparent magnitude    & LSST band brightness; higher redshift events are fainter and closer to selection limits & 0.045 \\
 3 & Hubble redshift ($z_{\rm HD}$) & fitted analysis quantity                                               & 0.035 \\
 4 & Heliocentric redshift          & fitted analysis quantity                                               & 0.030 \\
 5 & $r$ band apparent magnitude    & LSST band brightness                                               & 0.028 \\
 6 & classification consistency $\chi^2$  & analysis stage quantity              & 0.026 \\
 7 & \salt\ colour $c$              & Tripp standardisation quantity                                               & 0.021 \\
 8 & \salt\ stretch $x_1$           & Tripp standardisation quantity                                               & 0.020 \\
 9 & CMB frame redshift             & fitted analysis quantity                                               & 0.017 \\
10 & Peculiar velocity $\sigma_\mu$ & uncertainty quantity                                               & 0.017 \\
11 & bias correction covariance scale & analysis stage quantity                                         & 0.012 \\
12 & renormalised distance uncertainty  & ablation shows negligible impact   & 0.006 \\
13 & $i$ band apparent magnitude    & LSST band brightness                                               & 0.006 \\
14 & distance bias correction       & analysis stage quantity                                               & 0.006 \\
15 & $g$ band apparent magnitude    & LSST band brightness                                               & 0.005 \\
\enddata
\end{deluxetable*}

We examine the \(Y\) band apparent magnitude in more detail because it is the top ranked SHAP feature, not because we assume the \(Y\) filter is the physical cause. In the M25 simulation, \(Y\) band peak magnitude is strongly correlated with the fitted apparent magnitude \(m_B\) (\(r=0.993\)), peak signal to noise ratio (\(r=-0.889\)), and Hubble redshift \(z_{\rm HD}\) (\(r=0.806\)). These correlations are computed directly from the M25 catalogue as a check on what the high SHAP rank is tracing. Thus, in this model, \(Y\) band magnitude works as a compact tracer of several coupled quantities: brightness, distance, colour/dust, signal to noise, and proximity to the selection limits. Other observer frame band magnitudes carry related information, but \(Y\) ranks highest. This ranking is cadence dependent and reflects the relative filter depths of the simulated sample (which is heavily DDF dominated); other survey cadences may rank other bands higher.

This interpretation is deliberately predictive rather than causal. The high SHAP rank does not imply a unique \(Y\) band calibration error, nor does it mean that equivalent diagnostics in other surveys must be expressed through an LSST \(Y\) band. It means that, for the M25 feature set and redshift distribution, the model finds \(Y\) band magnitude to be the most compact observable summary of the residual structure. Establishing the physical root cause would require controlled one at a time simulation changes, such as changing the dust model, selection model, or band calibration while holding the rest of the pipeline fixed.

\begin{figure*}[!tp]
\centering
\includegraphics[width=\textwidth]{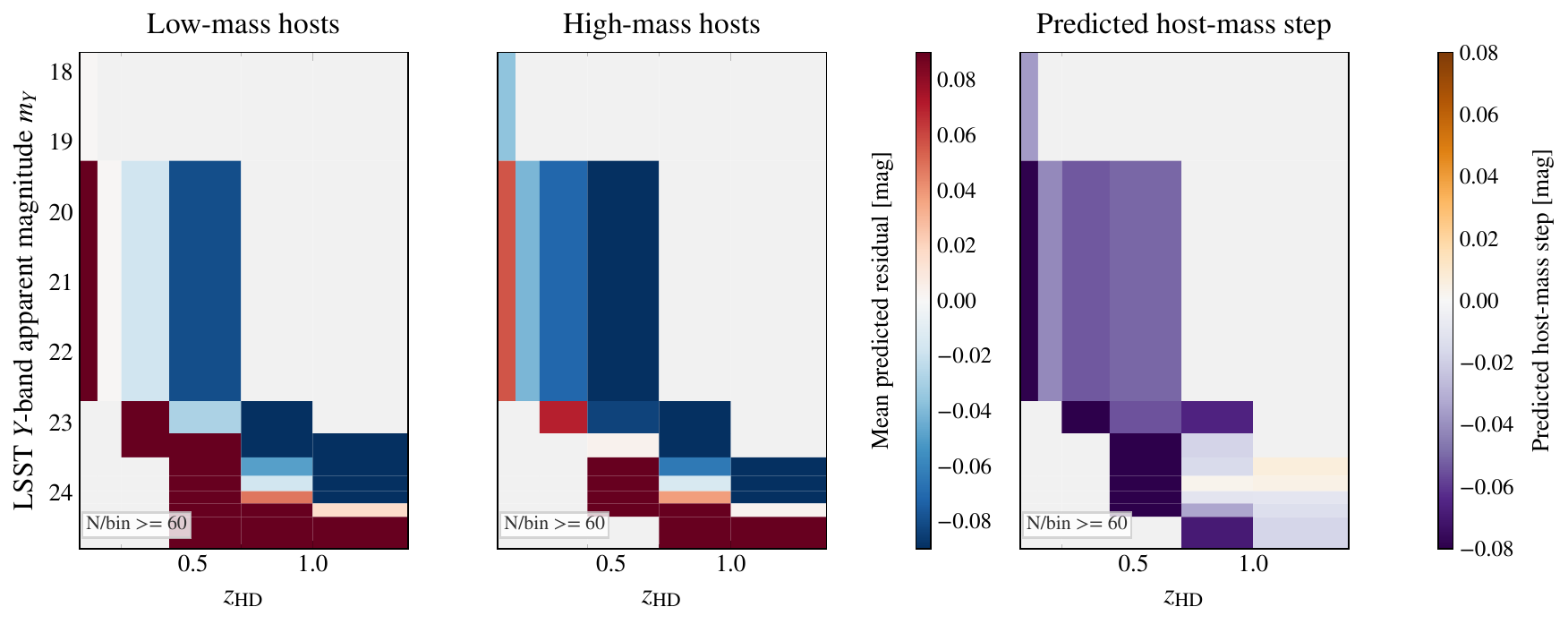}
\caption{Observed manifold dependence of the full M25 residual model. Each coloured cell contains the mean held out model prediction \(\widehat{\Delta\mu}\) for simulated Type Ia events in that populated \(z_{\rm HD}\) and LSST \(Y\) band apparent magnitude bin. The left panel uses low mass hosts (\(\logm<10\)); the middle panel uses high mass hosts (\(\logm\geq10\)). In those panels, red means positive predicted \(\Delta\mu\), blue means negative predicted \(\Delta\mu\), and the colour bar gives the value in mag. The right panel subtracts the low mass cell mean prediction from the high mass cell mean prediction at the same redshift and \(Y\) band magnitude. Purple therefore denotes a negative predicted high minus low host residual and orange denotes a positive one. Blank grey cells contain fewer than 60 events in either host bin and are omitted. The right panel is a conditional map at fixed redshift and apparent magnitude, not a marginalised host step or a causal partial dependence curve.\label{fig:model_dependence_surface}}
\end{figure*}

Figure~\ref{fig:model_dependence_surface} provides a complementary map of what the model learns on the observed simulation support. The x axis is Hubble redshift \(z_{\rm HD}\), the y axis is LSST \(Y\) band peak apparent magnitude, and the sample is split by recovered host galaxy stellar mass at \(\log(M_*/M_\odot)=10\). The model is first trained to predict the measured residual \(\Delta\mu\) for held out events. We then average those held out predictions, not the measured residuals themselves, inside each populated redshift magnitude host bin. The left and middle panels show the mean prediction, \(\widehat{\Delta\mu}\), for low  and high mass hosts. The right panel subtracts the low mass mean prediction from the high mass mean prediction in the same \(z_{\rm HD}\) \(Y\) magnitude cell. This difference is the model predicted host mass step at fixed redshift and apparent magnitude. Negative values mean that high mass hosts are predicted to have more negative residuals than low mass hosts in that cell.

The map shows that the predicted host mass step is not a single constant offset. It varies across the observed redshift brightness plane, with the largest negative values around intermediate redshift where selection effects are strong. Because the figure averages only over cells populated by M25 events, it should be read as an observed manifold summary of the fitted model, not as a causal partial dependence plot.

To check that the feature importance interpretation is not a single fold artefact, we compute LightGBM feature rankings separately on each of the five cross validation folds. The ranking is highly stable, with mean Spearman rank correlation $\langle\rho\rangle = 0.990$ and top 20 Jaccard overlap \(J = 0.933\). Reduced models without bias correction, error, or per band photometry features show similarly stable rankings.

\subsection{Feature Family Ablations}
\label{sec:ablations}

\begin{figure*}[!tp]
\centering
\includegraphics[width=0.82\textwidth]{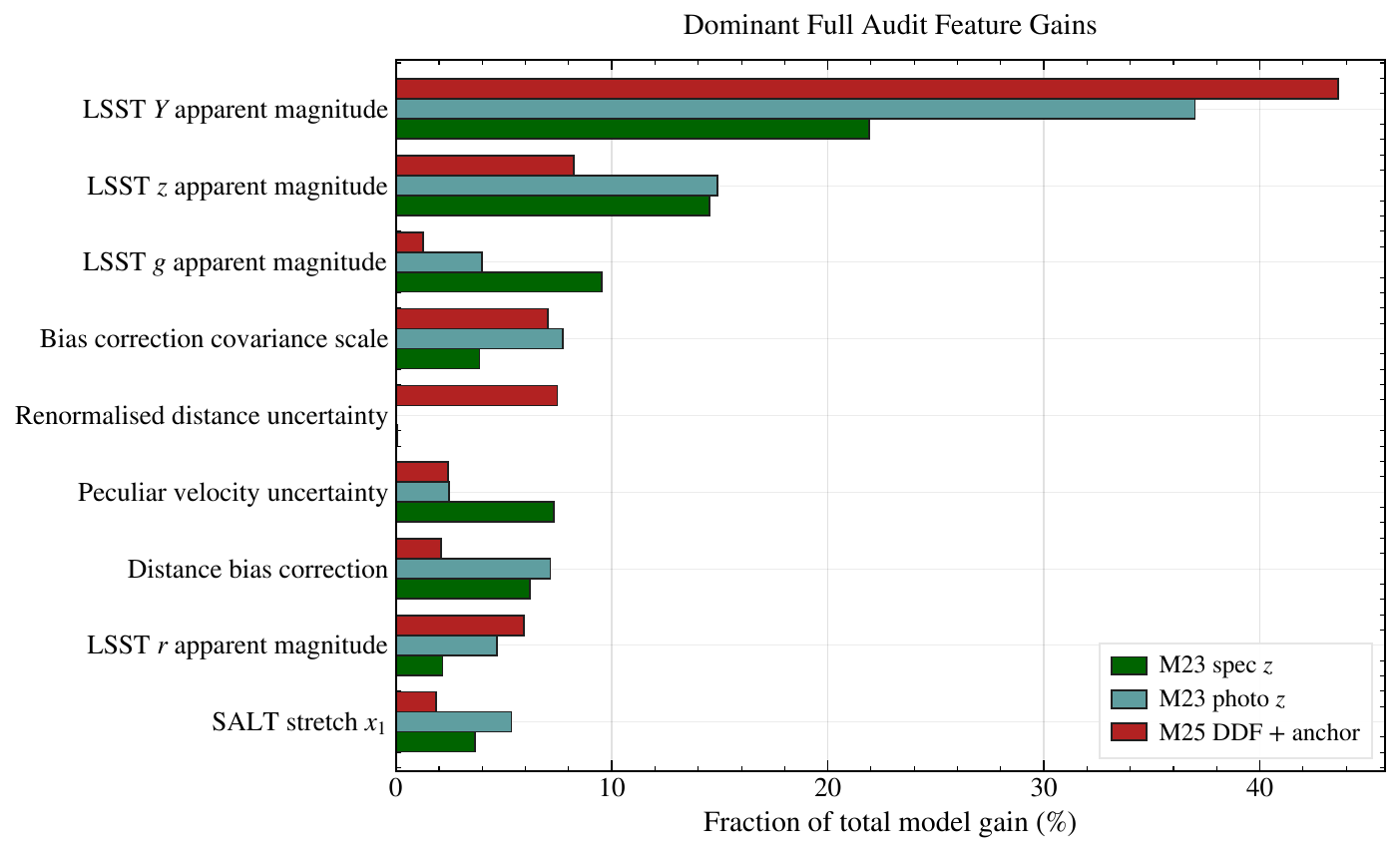}
\caption{Feature gains, shown as the fraction of the total full model gain beyond the Tripp standardisation model that is associated with each feature family. A larger bar indicates that the corresponding feature family contains predictive information not captured by the standardisation variables alone. Observer frame band specific apparent magnitudes, such as \(Y\) band magnitude, are raw photometric observables and are distinct from the fitted rest frame Tripp parameter \(m_B\).\label{fig:lsst_feature_gain}}
\end{figure*}

\begin{figure*}[!tp]
\centering
\includegraphics[width=0.82\textwidth]{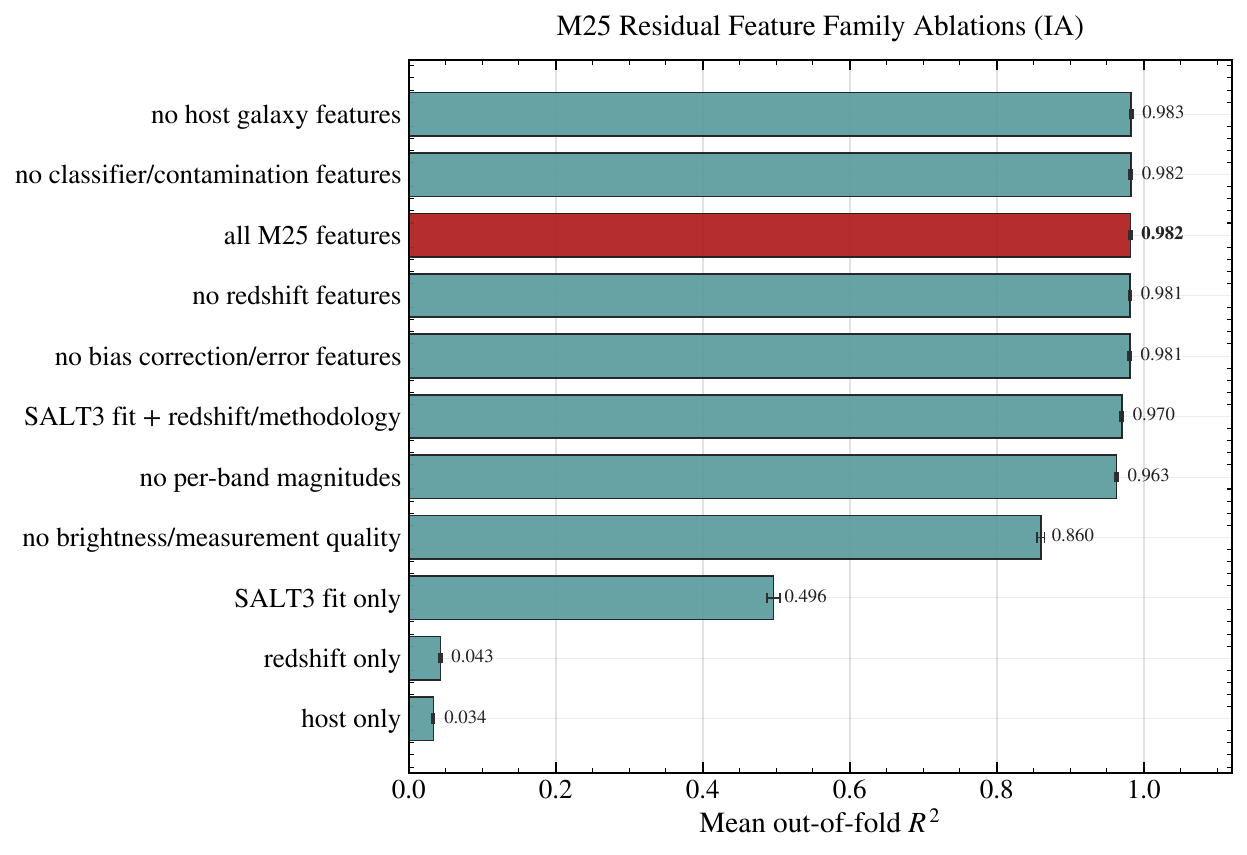}
\caption{Out of fold $R^2$ performance of the \lgbm\ model under feature family ablations. Removing the brightness and measurement quality family causes the largest drop in predictive performance. Redshift only and host only models have \(R^2 \approx 0\), showing that these variables do not dominate the residuals as isolated one dimensional predictors. SHAP and ablation answer different questions: SHAP ranks contributions inside the full model, whereas ablation measures how well the model can recover after a feature family is removed.\label{fig:ablation}}
\end{figure*}

To identify which groups of features carry the residual predictability, we train separate \lgbm\ models after removing specific feature families (Figures~\ref{fig:lsst_feature_gain} and \ref{fig:ablation}).

The combined light curve plus redshift model performs well, whereas models using only light curve fits or only redshifts perform poorly. This is expected because the target is a distance residual: the model needs both a brightness scale from the light curve fit and a distance scale from redshift to reconstruct a distance like quantity.

Three key results emerge:
\begin{enumerate}
  \item \textbf{Brightness and measurement quality dominate.} This family contains the fitted Tripp apparent magnitude \(m_B\), the peak signal to noise ratios in all observer frame bands, and the observer frame peak apparent magnitudes. The name does not mean that \(m_B\) is an SNR feature; rather, the family groups brightness scale and measurement quality observables. Removing the full family drops \(R^2\) from 0.982 to 0.860, the largest single ablation effect. Removing only the per band peak apparent magnitudes gives \(R^2=0.963\) $\implies$ \(\Delta R^2=0.982-0.963=0.019\).

  \item \textbf{Explicit redshift is not required for prediction.} Removing all explicit redshift columns, such as Hubble diagram, CMB frame, and heliocentric redshifts, changes \(R^2\) by only \(-0.001\). This does not mean that redshift is physically absent from the prediction. Rather, in the populated M25 sample, observer frame apparent magnitudes and signal to noise ratios are strongly correlated with redshift because more distant events are fainter and closer to the flux limit. These observables therefore provide empirical proxies for the same distance selection ordering, mixed with colour, dust, and measurement quality information.

  \item \textbf{Host galaxy and bias correction features are weak in isolation.} A model using only host galaxy features (\(R^2=0.034\)) or only bias correction/error features cannot predict most of the residual variance. The host mass signal is therefore learned as part of a broader correlated observable space rather than as a dominant one dimensional host effect.
\end{enumerate}

This also explains why Figure~\ref{fig:ablation} and the SHAP ranking are consistent. A high SHAP value for \(Y\) band magnitude means that the full model often uses that feature. A modest ablation drop after removing only per band magnitudes means that other correlated features, such as \(m_B\), redshift proxies, and signal to noise ratios, can partly substitute for them. In a closure audit, low \(R^2\) is the desired physical outcome; in an ablation, the drop in \(R^2\) tells us which removed features had been carrying predictive information.

\subsection{Standalone Real Data Application: DES 5 Year Audit}
\label{sec:des_audit}

To demonstrate that the closure audit can be applied to real observations, we run the same diagnostic directly on the Dark Energy Survey 5 Year (DES 5YR) spectroscopic Type Ia supernova sample from the DES SN5YR analysis \citep{Vincenzi2024}. This data set contains 1,820 spectroscopically confirmed SNe~Ia after the feature matching used here, analysed with SALT3 standardisation and simulation based bias corrections.

We train and evaluate the \lgbm\ model on the DES 5YR sample using five fold cross validation and the same hyperparameters. For the M25 DES comparison, we intentionally restrict both models to the 13 features available in both catalogues: light curve stretch ($x_1$), colour ($c$), apparent magnitude ($m_B$), light curve fit quality and fit probability, host galaxy properties (reconstructed host stellar mass $\logm_{\rm obs}$, directional light radius distance $d_{\rm DLR}$, and number of matching candidates), Milky Way reddening, and peak signal to noise ratio in three ordered bands. This reduced feature set is not meant to be the optimal DES only audit. It is a controlled common feature comparison. A survey specific DES audit should retain DES \(griz\) band information. We omit DES only band magnitudes here only because they would make the M25 and DES feature spaces different and would turn the comparison into a survey specific DES audit rather than a matched shared feature test.

\subsubsection{Audit Metrics on Real Data}

Table~\ref{tab:predictability_performance} compares the cross validation performance of the M25 reduced shared feature model and the DES shared feature model. The M25 model achieves \(R^2 = 0.967\) and RMSE \(=0.041\)~mag on simulated data, while the DES model achieves \(R^2 = 0.725\) and RMSE \(=0.139\)~mag on real data.

The DES result shows that observable dependent residual predictability is present in real data products as well as in simulations. The lower \(R^2\) and larger RMSE in DES are expected because real observations include larger intrinsic scatter, heterogeneous observing conditions, and unmodelled systematics that are not present in the simplified simulation universe. The fact that using 13 features alone already captures $R^2 = 0.725$, we didnot move further to test $R^2$ values for full DES feature set. 

\subsubsection{Feature Attribution Comparison}

\begin{figure*}[!tp]
\centering
\includegraphics[width=0.82\textwidth]{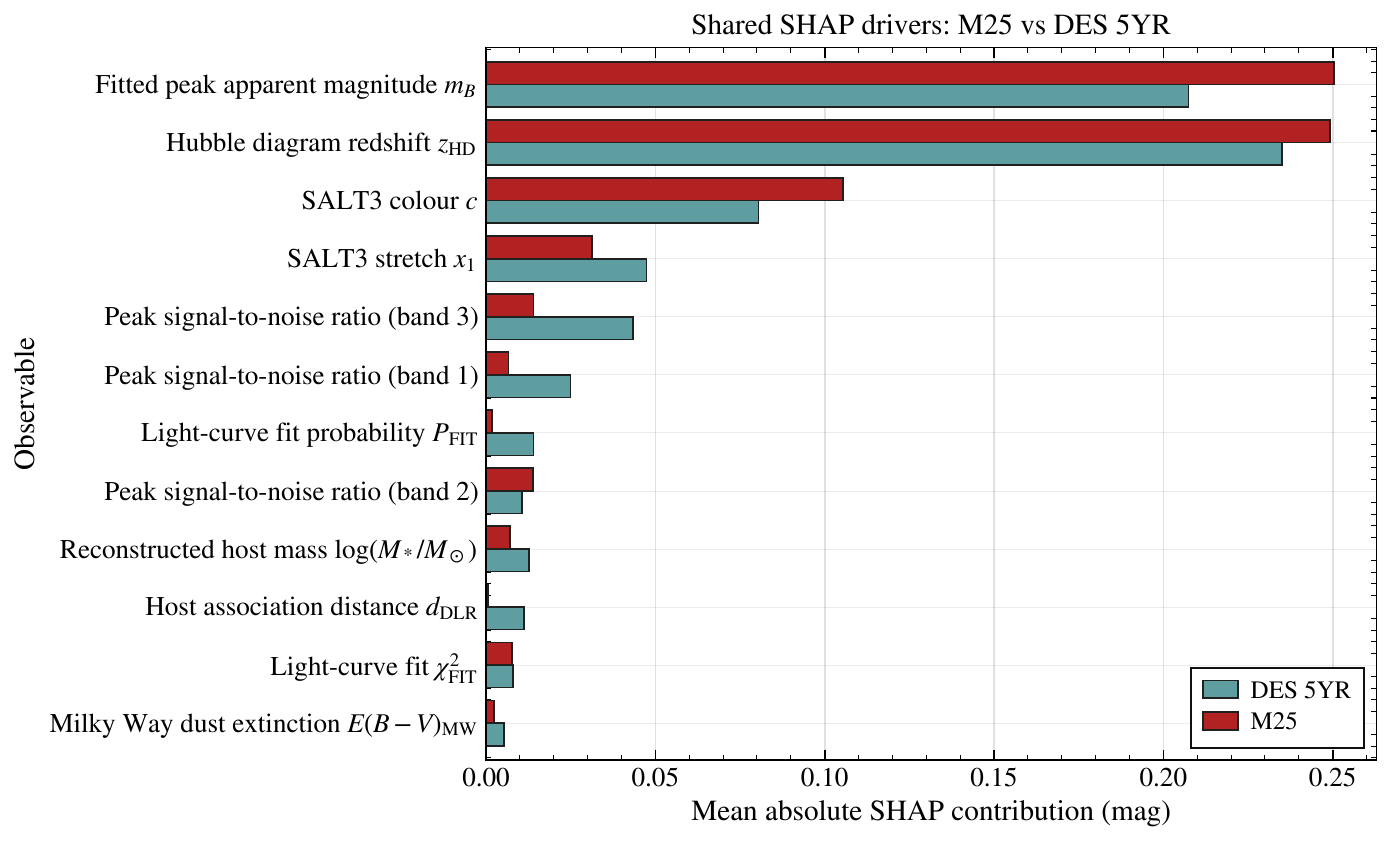}
\caption{Comparison of mean absolute SHAP feature attributions between the M25 shared feature model (blue) and the DES shared feature model (orange). The x-axis is mean absolute SHAP contribution in mag in both cases. The feature importances are highly correlated ($\rho = 0.802$), with apparent magnitude ($m_B$), redshift ($z_{\rm HD}$), colour ($c$), and stretch ($x_1$) dominating in both models. The agreement shows that the two models use a similar predictive feature hierarchy in the shared feature space; it does not establish a unique physical cause.\label{fig:des_native_shap}}
\end{figure*}

To test whether the predictable structure in real observations is associated with the same broad feature hierarchy as in the simulation, we compare mean absolute SHAP values from the M25 and DES shared feature models. Table~\ref{tab:des_vs_m25_shap_rank} gives the feature importances and rank orders, and Figure~\ref{fig:des_native_shap} compares the same values as a bar chart. In both data sets, apparent magnitude \(m_B\) and redshift \(z_{\rm HD}\) dominate, followed by light curve colour \(c\) and stretch \(x_1\). The Spearman rank correlation between M25 and DES feature importances is \(\rho = 0.802\).

Even when restricted to observables common to both data products, the variables that predict residuals are ranked similarly. This correlation suggests that the simulation captures the broad hierarchy of observable coordinates used to predict residuals in DES. We do not interpret the agreement as proof that the same physical mechanism drives both data sets; it is evidence that the audit is comparing similar predictive structure in the shared feature space.

\begin{table*}[t]
\centering
\caption{Feature SHAP importances and ranks for the M25 and DES shared feature models. The columns are: (1) \textbf{Observable}: the parameters shared between the simulation and real data catalogues; (2) \textbf{M25}: the mean absolute SHAP value in mag and rank in the simulation model; (3) \textbf{DES 5YR}: the mean absolute SHAP value in mag and rank in the real data model; and (4) \textbf{Rank Delta (DES $-$ M25)}: the change in rank order between data sets. A low rank delta  indicate similar predictive feature hierarchy in the shared feature space, not proof of a unique physical cause. Features are ordered by their M25 rank.\label{tab:des_vs_m25_shap_rank}}
\begin{tabular}{lccccc}
\hline\hline
Observable & \multicolumn{2}{c}{M25 simulation} & \multicolumn{2}{c}{DES 5YR real data} & Rank Delta \\
 & Mean $|\mathrm{SHAP}|$ (mag) & Rank & Mean $|\mathrm{SHAP}|$ (mag) & Rank & (DES $-$ M25) \\
\hline
Peak apparent magnitude $m_B$           & 0.250 & 1  & 0.207 & 2  & +1  \\
Hubble diagram redshift $z_{\rm HD}$    & 0.249 & 2  & 0.235 & 1  & $-1$ \\
SALT3 colour $c$                         & 0.105 & 3  & 0.080 & 3  & 0   \\
SALT3 stretch $x_1$                      & 0.031 & 4  & 0.047 & 4  & 0   \\
Peak signal to noise (band 3)            & 0.014 & 5  & 0.043 & 5  & 0   \\
Peak signal to noise (band 2)            & 0.014 & 6  & 0.011 & 10 & +4  \\
Light curve fit $\chi^2_{\rm FIT}$       & 0.008 & 7  & 0.008 & 11 & +4  \\
Reconstructed host mass $\logm_{\rm obs}$ & 0.007 & 8  & 0.013 & 8  & 0   \\
Peak signal to noise (band 1)            & 0.007 & 9  & 0.025 & 6  & $-3$ \\
Milky Way dust extinction $E(B-V)_{\rm MW}$ & 0.002 & 10 & 0.005 & 12 & +2  \\
Light curve fit probability $P_{\rm FIT}$ & 0.002 & 11 & 0.014 & 7  & $-4$ \\
Host association distance $d_{\rm DLR}$  & 0.001 & 12 & 0.011 & 9  & $-3$ \\
Host association candidates $N_{\rm MATCH}$ & 0.000 & 13 & 0.000 & 13 & 0   \\
\hline
\end{tabular}
\end{table*}

For heterogeneous compilations such as Pantheon or Pantheon+ \citep{Scolnic2018, Brout2022}, the audit should be applied at two levels. First, each survey component should have its own scorecard against its corresponding simulation, preserving survey specific observables such as band specific signal to noise ratios. Second, a compilation wide scorecard can be run on a deliberately restricted common feature set, for example SALT fit parameters, redshift, host stellar mass, and a categorical Survey ID. In a closed compilation, Survey ID should not be a strong predictor of residuals. If it becomes important, it flags a calibration, selection, or simulation domain mismatch between survey components. The DES 5YR spectroscopic sample is useful here because it is comparatively homogeneous and processed through a single analysis pipeline.

\subsection{Robustness to Photometric Calibration Systematics}
\label{sec:calibration_systematics}

We test the robustness of the closure audit by repeating the full \lgbm\ model training and evaluation across 19 M25 systematic analysis configurations (variants) available in the analysis products. These configurations represent the nominal analysis plus 18 non nominal systematic variants (commonly termed FITOPT configurations in the parent pipeline), each evaluated on all 25 M25 realisations. This is the complete set of variants used in this robustness audit, not an exhaustive list of every possible LSST systematic. The non nominal variants include per band zero point shifts (+5~mmag) in $ugrizY$, per band filter wavelength shifts (+5~\AA) in $ugrizY$, Milky Way dust extinction rescaling, spectroscopic redshift calibration, photometric redshift host galaxy shift, and alternative selection cuts. Appendix~\ref{sec:appendix_fitopt_inventory} lists the exact configuration labels and perturbation arguments.

The model performance is stable across these systematics: the out of fold \(R^2\) remains near \(0.981\), with degradation to \(0.965\) and \(0.969\) for the \(g\) band zero point and wavelength shifts, respectively; the RMSE remains near \(0.031\)~mag; and the feature rankings remain consistent with the nominal model (Spearman \(\rho \geq 0.9909\) across all variants).

The two largest degradations occur for the \(g\) band zero point and wavelength shifts. These perturbations most directly affect rest frame blue colour \(c\) and peak apparent magnitude \(m_B\) at low redshift, where LSST \(g\) samples the rest frame SED peak. The SHAP hierarchy nevertheless remains stable, indicating that the learned residual structure is an empirical feature of the analysis pipeline rather than an artefact of one calibration variant.

\section{Discussion}
\label{sec:discussion}

The closure audit shows that post standardisation supernova cosmology residuals can be far from featureless noise, even when one dimensional redshift summaries appear closed. Some predictability is expected because the target residual contains fitted light curve and redshift terms. The diagnostic result is that held out models find coherent residual structure beyond simple redshift binned baselines and that the same broad feature hierarchy appears in both M25 and DES. The scorecard is therefore a triage tool: it identifies stable observable dependent residual structures that warrant targeted modelling checks before cosmological unblinding.

Using simulations to audit residual fields is standard in other high stakes domains. In high energy physics, the ATLAS and CMS collaborations at the Large Hadron Collider rely on data to simulation closure tests to verify detector calibrations and assign systematic uncertainties \citep{ATLAS2021}. Similarly, deep learning medical image reconstruction analyses residual error fields between network outputs and physical phantoms to prevent artefacts and ensure clinical accuracy \citep{Antun2020, Colbrook2022}. In both contexts, residual structure is treated as a diagnostic warning sign; our scorecard brings the same logic to supernova cosmology.

Operating this framework on future LSST data sets requires pre survey reference scorecards trained on approved simulation configurations (Table~\ref{tab:scorecard_template}). When real LSST observations arrive, the same scorecard can be executed independently on the corrected data and on the mock ensemble. We propose a simple operational flag: if a real data metric differs from the mock ensemble mean by more than twice the mock to mock scatter, \(>2\sigma_{\rm mock}\), the case is treated as a potential closure failure or domain gap. Here \(\sigma_{\rm mock}\) is the standard deviation of the same scorecard metric across matched mock realisations. The empirical tail probability is the fraction of mock realisations whose metric is at least as far from the mock mean as the real data value. The motivation is pragmatic rather than fundamental: for an approximately Gaussian mock distribution, a two sided \(2\sigma\) discrepancy corresponds to a tail probability of about 0.05, and a finite mock ensemble can report this tail probability directly. A flag should trigger controlled one at a time simulation checks, such as alternate dust scatter or host mass evolution models, rather than an immediate machine learning correction to the Hubble diagram. The numerical thresholds in Table~\ref{tab:scorecard_template} are therefore proposed operational starting points to be calibrated by survey specific cosmology impact studies, not universal physical constants.

\begin{table*}[t]
\centering
\footnotesize
\caption{Standardised closure audit scorecard template and diagnostic checklist. The columns are: (1) \textbf{Audit Category}: the diagnostic family; (2) \textbf{Target Metric}: the monitored statistic, such as out of fold \(R^2\), step differences, or classification AUC; (3) \textbf{Features Used}: the observable subset given to the model; (4) \textbf{Normal Expectation}: the baseline expected under closure; and (5) \textbf{Trigger/Action for Non Closure}: proposed thresholds and follow up actions. The thresholds are operational flags to be calibrated with survey specific cosmology impact studies.\label{tab:scorecard_template}}
\begin{tabular}{lllll}
\tableline\tableline
Audit Category & \shortstack[l]{Target\\Metric} & \shortstack[l]{Features\\Used} & \shortstack[l]{Normal\\Expectation} & \shortstack[l]{Trigger/Action for Non Closure} \\
\tableline
\textbf{1. Predictability Audits} \\
~~Global Audit & \shortstack[l]{Out of fold $R^2$\\(e.g., LightGBM)} & \shortstack[l]{Full\\feature set} & $R^2 \approx 0$ & \shortstack[l]{$R^2 \ge 0.10$: unmodelled residual\\structure. Run controlled sub sims.} \\
~~Ablation Gap & \shortstack[l]{$R^2_{\rm full} - R^2_{\rm core}$} & \shortstack[l]{Full vs.\\core parameters} & $\Delta R^2 \approx 0$ & \shortstack[l]{$\Delta R^2 \ge 0.05$: Multivariate\\selection bias leakage.\\Inspect band specific SNR/flux\\and selection cuts.} \\
~~Redshift Residuals & \shortstack[l]{Out of fold $R^2_z$} & Redshift only & $R^2_z \approx 0$ & \shortstack[l]{$R^2_z \ge 0.02$: Incomplete global/redshift \\binned distance bias correction.} \\
\hline
\textbf{2. Target Systematic Audits} \\
~~Host Mass Step & \shortstack[l]{Observed vs.\\predicted step\\($\gamma_{\rm obs} - \gamma_{\rm pred}$)} & \shortstack[l]{Host stellar\\mass} & \shortstack[l]{$\gamma_{\rm obs} -$\\ $\gamma_{\rm pred} \approx 0$} & \shortstack[l]{Discrepancy $\ge 0.01$~mag: Host mass\\dust/metallicity evolution mismatch.} \\
~~Redshift Step Profile & \shortstack[l]{Bin by bin\\mass step $\gamma(z)$} & \shortstack[l]{Host stellar\\mass} & \shortstack[l]{Stable/flat\\across $z$} & \shortstack[l]{Reversal/evolution: Mass dependent\\selection bias or dust model failure.} \\
\hline
\textbf{3. Domain Gap Audits} \\
~~Adversarial Classifier & \shortstack[l]{Domain classifier\\AUC or $R^2$} & \shortstack[l]{All\\observables} & \shortstack[l]{${\rm AUC} \approx 0.50$\\(indistinguishable)} & \shortstack[l]{${\rm AUC} \ge 0.60$: Domain gap. \\Mock physics is incomplete.} \\
~~Real Data Outlier & \shortstack[l]{Real scorecard\\position in\\mock distribution} & \shortstack[l]{Real vs.\\mock ensemble} & \shortstack[l]{Within $2\sigma$ of\\mock ensemble} & \shortstack[l]{Out of distribution real score.\\Refine mock simulation parameters\\before unblinding.} \\
\tableline
\end{tabular}
\end{table*}

Several caveats limit the scope of this diagnostic. First, the closure audit is a relative validation tool: if a systematic error is absent from both the data and bias correction simulations, the pipeline may falsely close. Second, the scorecard does not directly quantify the induced bias in cosmological parameters such as \(w\) or \(\Omega_m\); any proposed mitigation must be propagated through a fully blinded end to end cosmological analysis. Appendix~\ref{sec:appendix_cosmo} shows why applying the ML model as a direct correction is unsafe and cosmology dependent. Finally, the framework is not restricted to SALT or BBC, and can be applied to residuals from hierarchical Bayesian estimators such as BayeSN \citep{Grayling2024} or simulation based models like FlowSN \citep{Boyd2026}.

\section{Conclusions}
\label{sec:conclusions}

We introduce a machine learning closure audit for simulation dependent supernova cosmology pipelines. Our principal conclusions are:
\begin{enumerate}
  \item Audits of M23 and M25 reveal substantial conditional residual structure; one dimensional redshift bin means explain \(<1\%\) of the variance, while full out of fold models reach \(R^2 \ge 0.94\).
  \item The scorecard combines gains beyond simple baselines with stable conditional structures, null tests, and robustness under systematic variations.
  \item The direct audit of the real DES 5YR spectroscopic sample achieves \(R^2 = 0.725\) and RMSE \(= 0.139\)~mag, demonstrating that predictable residuals exist in real telescope data.
  \item The SHAP feature attribution rankings are consistent between independent M25 and DES shared feature models (Spearman rank correlation \(\rho = 0.802\)), dominated by apparent magnitude, redshift, colour, and stretch. This indicates similar predictive structure in the shared feature space.
  \item For future real LSST data, the scorecard can be executed independently on corrected real data and mock ensembles to identify systematic domain gaps before cosmological unblinding.
\end{enumerate}

The operational conclusion is that a simulation can recover acceptable global summaries while retaining scientifically important conditional residual structure. The closure audit scorecard makes that structure measurable before real LSST data exist and provides a defined comparison to perform when they do.

\appendix
\section{Audit Feature Inventories}
\label{sec:appendix_feature_inventory}

Tables~\ref{tab:audit_feature_inventory_fit}--\ref{tab:audit_feature_inventory_analysis} list the exact input observables used by the reported feature-space audits. The inventories are derived from the saved model manifests, after identifier, simulation-truth, direct target-leakage, unavailable, and mostly missing columns have been removed.

\startlongtable
\begin{deluxetable*}{llccc}
\tabletypesize{\scriptsize}
\tablecaption{Audit feature inventory, light-curve, redshift, foreground, and lensing observables.}
\label{tab:audit_feature_inventory_fit}
\tablewidth{0pt}
\tablehead{
  \colhead{Group} & \colhead{Feature} & \colhead{M25 full} & \colhead{M23 full} & \colhead{DES shared}
}
\startdata
Light-curve fit & FITCHI2 & yes & yes & yes \\
Light-curve fit & FITPROB & yes & yes & yes \\
Light-curve fit & NDOF & yes & yes & \nodata \\
Light-curve fit & c & yes & yes & yes \\
Light-curve fit & cERR & yes & yes & \nodata \\
Light-curve fit & mB & yes & yes & yes \\
Light-curve fit & mBERR & yes & yes & \nodata \\
Light-curve fit & x1 & yes & yes & yes \\
Light-curve fit & x1ERR & yes & yes & \nodata \\
Redshift and velocity & HOST\_QZPHOT & yes & \nodata & \nodata \\
Redshift and velocity & HOST\_QZPHOTSTD & yes & \nodata & \nodata \\
Redshift and velocity & HOST\_ZPHOT & yes & yes & \nodata \\
Redshift and velocity & HOST\_ZPHOTERR & yes & yes & \nodata \\
Redshift and velocity & OPT\_PHOTOZ & yes & yes & \nodata \\
Redshift and velocity & VPEC & yes & yes & \nodata \\
Redshift and velocity & VPECERR & yes & yes & \nodata \\
Redshift and velocity & zCMB & yes & yes & \nodata \\
Redshift and velocity & zCMBERR & yes & yes & \nodata \\
Redshift and velocity & zHD & yes & yes & yes \\
Redshift and velocity & zHDERR & yes & yes & \nodata \\
Redshift and velocity & zHEL & yes & yes & \nodata \\
Foreground/lensing & LENSDMU & yes & \nodata & \nodata \\
Foreground/lensing & LENSDMUERR & yes & \nodata & \nodata \\
Foreground/lensing & MWEBV & yes & yes & yes \\
\enddata
\tablecomments{Columns mark whether the observable enters the reported M25 76-feature full audit, the M23 58-feature full audit, or the 13-feature M25--DES shared audit. The M23 column applies to both the M23 spectroscopic-redshift and photometric-redshift samples.}
\end{deluxetable*}

\startlongtable
\begin{deluxetable*}{llccc}
\tabletypesize{\scriptsize}
\tablecaption{Audit feature inventory, host-galaxy observables.}
\label{tab:audit_feature_inventory_host}
\tablewidth{0pt}
\tablehead{
  \colhead{Group} & \colhead{Feature} & \colhead{M25 full} & \colhead{M23 full} & \colhead{DES shared}
}
\startdata
Host galaxy & HOST\_ANGSEP & yes & yes & \nodata \\
Host galaxy & HOST\_CONFUSION & yes & \nodata & \nodata \\
Host galaxy & HOST\_DDLR & yes & yes & yes \\
Host galaxy & HOST\_LOGMASS & yes & \nodata & yes \\
Host galaxy & HOST\_NMATCH & yes & yes & yes \\
Host galaxy & HOST\_NMATCH2 & yes & yes & \nodata \\
Host photometry & HOST\_MAG\_Y & yes & \nodata & \nodata \\
Host photometry & HOST\_MAG\_g & yes & \nodata & \nodata \\
Host photometry & HOST\_MAG\_i & yes & \nodata & \nodata \\
Host photometry & HOST\_MAG\_r & yes & \nodata & \nodata \\
Host photometry & HOST\_MAG\_u & yes & \nodata & \nodata \\
Host photometry & HOST\_MAG\_z & yes & \nodata & \nodata \\
Host photometry & HOST\_SBMAG\_Y & yes & yes & \nodata \\
Host photometry & HOST\_SBMAG\_g & yes & yes & \nodata \\
Host photometry & HOST\_SBMAG\_i & yes & yes & \nodata \\
Host photometry & HOST\_SBMAG\_r & yes & yes & \nodata \\
Host photometry & HOST\_SBMAG\_u & yes & yes & \nodata \\
Host photometry & HOST\_SBMAG\_z & yes & yes & \nodata \\
\enddata
\tablecomments{Columns mark whether the observable enters the reported M25 76-feature full audit, the M23 58-feature full audit, or the 13-feature M25--DES shared audit. The M23 column applies to both the M23 spectroscopic-redshift and photometric-redshift samples.}
\end{deluxetable*}

\startlongtable
\begin{deluxetable*}{llccc}
\tabletypesize{\scriptsize}
\tablecaption{Audit feature inventory, SN photometry, light-curve coverage, and analysis-stage outputs.}
\label{tab:audit_feature_inventory_analysis}
\tablewidth{0pt}
\tablehead{
  \colhead{Group} & \colhead{Feature} & \colhead{M25 full} & \colhead{M23 full} & \colhead{DES shared}
}
\startdata
SN photometry/coverage & PKMJD & yes & yes & \nodata \\
SN photometry/coverage & PKMJDERR & yes & yes & \nodata \\
SN photometry/coverage & SNRMAX1 & yes & yes & yes \\
SN photometry/coverage & SNRMAX2 & yes & yes & yes \\
SN photometry/coverage & SNRMAX3 & yes & yes & yes \\
SN photometry/coverage & SNRMAX\_Y & yes & yes & \nodata \\
SN photometry/coverage & SNRMAX\_g & yes & yes & \nodata \\
SN photometry/coverage & SNRMAX\_i & yes & yes & \nodata \\
SN photometry/coverage & SNRMAX\_r & yes & yes & \nodata \\
SN photometry/coverage & SNRMAX\_z & yes & yes & \nodata \\
SN photometry/coverage & SNRSUM & yes & \nodata & \nodata \\
SN photometry/coverage & T0GAPMAX & yes & yes & \nodata \\
SN photometry/coverage & TGAPMAX & yes & yes & \nodata \\
SN photometry/coverage & TrestMAX & yes & yes & \nodata \\
SN photometry/coverage & TrestMIN & yes & yes & \nodata \\
SN photometry/coverage & TrestRange & yes & yes & \nodata \\
SN photometry/coverage & m0obs\_Y & yes & yes & \nodata \\
SN photometry/coverage & m0obs\_g & yes & yes & \nodata \\
SN photometry/coverage & m0obs\_i & yes & yes & \nodata \\
SN photometry/coverage & m0obs\_r & yes & yes & \nodata \\
SN photometry/coverage & m0obs\_u & yes & yes & \nodata \\
SN photometry/coverage & m0obs\_z & yes & yes & \nodata \\
Analysis-stage output & CHI2\_BEAMS & yes & \nodata & \nodata \\
Analysis-stage output & MUERR & yes & yes & \nodata \\
Analysis-stage output & MUERR\_RAW & yes & yes & \nodata \\
Analysis-stage output & MUERR\_RENORM & yes & yes & \nodata \\
Analysis-stage output & MUERR\_VPEC & yes & yes & \nodata \\
Analysis-stage output & PROBCC\_BEAMS & yes & \nodata & \nodata \\
Analysis-stage output & PROB\_SCONE\_PREDICT\_LSST\_SIMDATA & yes & \nodata & \nodata \\
Analysis-stage output & RADIUS\_POP & yes & \nodata & \nodata \\
Analysis-stage output & biasCorErr\_mu & yes & yes & \nodata \\
Analysis-stage output & biasCor\_mu & yes & yes & \nodata \\
Analysis-stage output & biasCor\_muCOVSCALE & yes & yes & \nodata \\
Analysis-stage output & biasCor\_nevt & yes & \nodata & \nodata \\
\enddata
\tablecomments{Columns mark whether the observable enters the reported M25 76-feature full audit, the M23 58-feature full audit, or the 13-feature M25--DES shared audit. The M23 column applies to both the M23 spectroscopic-redshift and photometric-redshift samples.}
\end{deluxetable*}

\section{M25 Systematic Analysis Configurations}
\label{sec:appendix_fitopt_inventory}

The inventory below lists the specific nominal and alternative (non nominal) analysis configurations (systematic variants, often referred to as FITOPT configurations in the parent pipeline) audited here.

\startlongtable
\begin{deluxetable*}{clll}
\tabletypesize{\scriptsize}
\tablecaption{M25 systematic analysis variants used in the robustness scan.}
\label{tab:m25_fitopt_inventory}
\tablewidth{0pt}
\tablehead{
  \colhead{Variant ID} & \colhead{Name} & \colhead{Family} & \colhead{Perturbation argument}
}
\startdata
000 & nominal & Nominal & none \\
001 & MWEBV & Foreground & MWEBV\_SCALE 0.95 \\
002 & zshift & Redshift & REDSHIFT\_FINAL\_SHIFT 0.0001 \\
003 & ZERRSCALE & Redshift & PRIOR\_ZERRSCALE 1.2 \\
004 & NOREJECT\_test & Selection & PRIOR\_ZERRSCALE 0.8 \\
005 & Cal\_HST & Calibration & MAGOBS\_SHIFT\_ZP\_PARAMS 0 0.00714 0 \\
006 & Cal\_ZP\_u & Zero point & MAGOBS\_SHIFT\_ZP 'u .005' \\
007 & Cal\_ZP\_g & Zero point & MAGOBS\_SHIFT\_ZP 'g .005' \\
008 & Cal\_ZP\_r & Zero point & MAGOBS\_SHIFT\_ZP 'r .005' \\
009 & Cal\_ZP\_i & Zero point & MAGOBS\_SHIFT\_ZP 'i .005' \\
010 & Cal\_ZP\_z & Zero point & MAGOBS\_SHIFT\_ZP 'z .005' \\
011 & Cal\_ZP\_Y & Zero point & MAGOBS\_SHIFT\_ZP 'Y .005' \\
012 & Cal\_wave\_u & Wavelength & FILTER\_LAMSHIFT 'u 5' \\
013 & Cal\_wave\_g & Wavelength & FILTER\_LAMSHIFT 'g 5' \\
014 & Cal\_wave\_r & Wavelength & FILTER\_LAMSHIFT 'r 5' \\
015 & Cal\_wave\_i & Wavelength & FILTER\_LAMSHIFT 'i 5' \\
016 & Cal\_wave\_z & Wavelength & FILTER\_LAMSHIFT 'z 5' \\
017 & Cal\_wave\_Y & Wavelength & FILTER\_LAMSHIFT 'Y 5' \\
018 & Photo\_shift & Host photo-z & HOSTGAL\_PHOTOZ\_SHIFT 0.01 \\
\enddata
\tablecomments{Description of the nominal and 18 non-nominal systematic analysis variants (often termed FITOPT configurations in the parent pipeline) evaluated for the M25 Type Ia robustness scan. Zero-point perturbations are +0.005 mag in the named band; filter wavelength shifts are +5 Angstrom in the named band.}
\end{deluxetable*}

\section{Cosmological Parameter Bias and ML Correction Safety}
\label{sec:appendix_cosmo}

To empirically calibrate diagnostic scorecard trigger thresholds, such as \(R^2 \ge 0.10\) or \(\Delta R^2 \ge 0.05\), and test the safety of applying machine learning as a cosmological correction, we conduct a controlled cosmology stress test. This test uses a supplementary multi cosmology ensemble of simulated Type Ia supernova Hubble diagrams generated under two distinct injected cosmological models:
\begin{itemize}
    \item \textbf{Model A (Target):} Injected cosmological parameters $w_0 = -0.8$, $\Omega_m = 0.35$.
    \item \textbf{Model B (Source):} Injected cosmological parameters $w_0 = -1.2$, $\Omega_m = 0.25$.
\end{itemize}
Both simulated ensembles are processed through standard light curve fitting, selection cuts, and one dimensional redshift binned bias corrections, yielding binned Hubble diagram products. Because host masses are absent in this supplementary simulation, we restrict the machine learning features to the four Tripp standardisation variables \(\{x_1, c, m_B, z_{\rm HD}\}\) and train a LightGBM regressor on binned training fold data to predict binned Hubble residuals.

We run two testing scenarios:
\begin{enumerate}
    \item \textbf{Within Cosmology Correction (Control):} The model is trained and applied to correct data within the same injected cosmology (e.g., training and correcting Model A).
    \item \textbf{Cross Cosmology Correction (Stress Test):} The model is trained on one cosmology (e.g., Model B) but applied to correct the Hubble diagram of the other cosmology (e.g., Model A).
\end{enumerate}
For all cases, we run a full cosmology fit using standard distance estimation software, with both a Planck CMB prior and a no CMB baseline, across all realisations. Table~\ref{tab:cosmo_bias_results} reports the fitted cosmological parameters before and after applying the ML predicted residual correction.

\begin{table*}[t]
\centering
\begingroup
\scriptsize
\setlength{\tabcolsep}{3pt}
\caption{Fitted cosmological parameters before and after applying ML residual corrections under within cosmology and cross cosmology configurations.}
\label{tab:cosmo_bias_results}
\begin{tabular}{llcccccc}
\tableline\tableline
Scenario & Correction State & Injected $w_0$ & Fitted $w_0$ & $\Delta w_0$ & Injected $\Omega_m$ & Fitted $\Omega_m$ & $\Delta \Omega_m$ \\
\tableline
\textbf{1. Within Cosmology (Control)} \\
~~Model A ($w_0 = -0.8$) & Before correction & $-0.800$ & $-0.817 \pm 0.026$ & $-0.017$ & $0.350$ & $0.347 \pm 0.008$ & $-0.003$ \\
~~Model A ($w_0 = -0.8$) & After ML correction & $-0.800$ & $-0.804 \pm 0.017$ & $-0.004$ & $0.350$ & $0.351 \pm 0.006$ & $+0.001$ \\
~~Model B ($w_0 = -1.2$) & Before correction & $-1.200$ & $-1.191 \pm 0.027$ & $+0.009$ & $0.250$ & $0.255 \pm 0.006$ & $+0.005$ \\
~~Model B ($w_0 = -1.2$) & After ML correction & $-1.200$ & $-1.183 \pm 0.030$ & $+0.017$ & $0.250$ & $0.258 \pm 0.007$ & $+0.008$ \\
\hline
\textbf{2. Cross Cosmology (Stress Test)} \\
~~Model B $\to$ Model A & Before correction & $-0.800$ & $-0.813 \pm 0.024$ & $-0.013$ & $0.350$ & $0.349 \pm 0.009$ & $-0.001$ \\
~~Model B $\to$ Model A & After ML correction & $-0.800$ & $-1.138 \pm 0.027$ & $-0.338$ & $0.350$ & $0.260 \pm 0.007$ & $-0.090$ \\
~~Model A $\to$ Model B & Before correction & $-1.200$ & $-1.186 \pm 0.031$ & $+0.014$ & $0.250$ & $0.254 \pm 0.008$ & $+0.004$ \\
~~Model A $\to$ Model B & After ML correction & $-1.200$ & $-0.829 \pm 0.022$ & $+0.371$ & $0.250$ & $0.350 \pm 0.008$ & $+0.100$ \\
\tableline
\end{tabular}

Values represent the mean and standard deviation across the simulated realisations, assuming a standard flat cosmological model fit with a Planck like CMB prior. All parameter shifts ($\Delta w_0, \Delta \Omega_m$) are measured relative to the true injected values of the target sample.
\endgroup
\end{table*}
The results demonstrate a crucial cosmology dependence in the machine learning correction:
\begin{itemize}
    \item When the training and target simulations share the same cosmology (Scenario 1), the correction is stable, reducing the raw distance bias and recovering the true cosmology with tight uncertainties.
    \item When the training and target cosmologies differ (Scenario 2), the machine learning model acts as a strong cosmological prior. Subtracted residuals pull the fitted parameters of the target sample toward the cosmology of the training sample, introducing severe systematic biases of \(\Delta w_0 \approx -0.338\) and \(+0.371\), and shifting \(\Omega_m\) by \(\sim\!0.09-0.10\).
\end{itemize}

This stress test provides empirical support for using the scorecard as a trigger rather than as a correction. Because the correction is highly sensitive to the assumed cosmology of the training mock, applying an ML correction directly to observations is unsafe and can introduce artificial cosmological parameter shifts. The closure scorecard should therefore be operated as a \emph{diagnostic triage instrument} that triggers physical modelling updates in the simulation, not as a direct correction to the cosmological data.

\clearpage 



\end{document}